\documentclass[showpacs,preprintnumbers,twocolumn,amsmath,amssymb]{revtex4}
\usepackage{graphicx}
\usepackage{dcolumn}
\usepackage{bm}

\begin{document}
%


\title{Gap symmetries from the neighbor coupling in square-lattice superconductors}
\author{W. LiMing} \email{wliming@scnu.edu.cn}
\affiliation{Dept. of Physics, and Institute for Condensed Matter Physics, School of
Physics and Telecommunication Engineering, South China Normal University, Guangzhou
510006, China} 
\date{\today}
\begin{abstract}
The gap symmetries of superconductivity are studied in this work. It is found that the
gap symmetries are simply determined by the 4-fold rotational symmetries of the coupling
potential on neighbor sites. A local on-site coupling potential results in the on-site
pairing with the conventional $s$-wave symmetry, but a coupling potential between the
nearest neighbors or the next-nearest neighbors results in the pairing on neighbor sites
with the $s^-$, $d_{x^2-y^2}$, $d_{xy}$, or $s_{x^2y^2}$ gap symmetries. It is proved
that both isotropic and anisotropic gap functions are allowed by the 4-fold rotational
symmetries of the coupling potential. Finally a numerical computation is performed to
demonstrate the gap symmetries. This neighbor coupling provides a unified picture for the
gap functions of the conventional and the high Tc superconductivity.
\end{abstract} 
 \maketitle

\section{Introduction}
A few decades of study on high-Tc superconductor confirmed the $d$-wave symmetry of the
gap function at least in cuprate superconductors, relative to the conventional $s$-wave
BCS gap symmetry\cite{Kotlier,Tsui}. The observation of the half quantum of magnetic flux
provides a direct detection to the d-wave gap symmetry in cuprates\cite{Tsui}. The
concept of $d$-wave order, however, is quite misleading to the physics of
superconductivity. In general, it is inferred from the antisymmetry requirement of the
wave function of a Cooper pair. Since electrons are fermions the wave function of a pair
of electrons must be antisymmetric under their exchange. When the spin part of the wave
function is antisymmetric the real space part should be symmetric. Then it was thought
that the angular momentum of a Cooper pair can only be even, {\it i.e.,} $L = 0,2,4,...$.
These quantum numbers led to the description of $s, d, g$-wave order parameters. This
picture is a good approximation for electrons on a atom, where the rotational isotropy of
space allows the angular momentum to be a good quantum number. When the coupling
potential between electrons is  attractive on one atom it can be indeed expected to
observe these gap symmetries. Nevertheless, for high-Tc superconductors, the coupling is
no longer on-site attractive, but is attractive only on neighbor sites. In this case the
space isotropy is broken by the crystal lattice and only the point group symmetries
remain. Therefore, the concept of $d$-wave symmetry is not valid for high Tc
superconductors\cite{yunping}, which in fact should contain all components of
$L=0,2,4,...$. Another argument comes from the newly discovered iron-based
superconductors, which have a gap symmetry $\sim \cos k_x \cos k_y$\cite{Yan,Parish}.
This symmetry cannot be explained by the above $s,d,g$-wave interpretation. It is called
an extended $s$-wave. Some researchers such as Yao {\it et al} claimed a $d_{xy}$ gap
symmetry on one hand and at the same work they have to admit a sign change for the gap on
$\Gamma$ and $M$\cite{Wang} in the Brillouin zone.  It is seen that the $s$- or $d$-wave
description becomes very controversial for the gap symmetries in iron-based
superconductors.

In this paper I argue that the gap symmetries of square lattice superconductors in fact
originate from the 4-fold rotational symmetries of the coupling potential on neighbor
sites. It has nothing to do with the angular momentum. I present a theoretical analysis
and a computational demonstration for this argument. It is shown that the attraction
between the nearest neighbors ($n.n.$) or the next nearest neighbors ($n.n.n.$) provides
a coupling for Cooper pairs, so that the gap functions of Cooper pairs take the
symmetries of the coupling potential between neighbors. It is interesting to notice that
most gap symmetries of square lattice superconductors can be reproduced in this
mechanism.

\section{Gap symmetries}
The BCS coupling Hamiltonian is written as
\begin{align}
H_{BCS} &= -V \sum_{kk'}
c^\dagger_{k\uparrow}c^\dagger_{-k\downarrow}c_{-k'\downarrow}c_{k'\uparrow}
\end{align}
Change it into the site configuration then one finds
\begin{align}
H_{BCS} &= -V \sum_{ij}
c^\dagger_{i\uparrow}c^\dagger_{i\downarrow}c_{j\downarrow}c_{j\uparrow}
\end{align}
It comes from an on-site coupling, $V({\bf r}_1, {\bf r}_2)=V\delta({\bf r}_1- {\bf
r}_2)$\cite{Annett1}. This coupling indicates that in fact a Cooper pairing can only
occur on the same site. The electrons in cuprate superconductors, however, have strong
on-site repulsion due to the strong correlation as seen in the Hubbard model. One usually
rules out the double occupancy through the method of Gutzwiller prejection\cite{Claudio}.
Because of this strong correlation effect the conventional BCS on-site pairing is
certainly not energetically preferential in the cuprate superconductors. Hence, the
interaction Hamiltonian has to be changed to the coupling between neighbor sites as
follows
\begin{align}
H_{int} &= {1\over N} \sum_{ijm} V_m
c^\dagger_{i+m\uparrow}c^\dagger_{i\downarrow}c_{j\downarrow}c_{j+m\uparrow}
\end{align}
where $V_m$ is the coupling potential between two electrons (or holes) on two neighbor
sites, here $m$ denote the neighbors of a fixed site. In general, only the potential
values on the near neighbors, such as $V_0,V_1, V_2$, are important.
 In the mean field approximation the
above Hamiltonian becomes
\begin{align}
H_{int} &= -\sum_{k} (\Delta^*_k c_{-k\downarrow}c_{k\uparrow} + \Delta_k
c^\dagger_{k\uparrow}c^\dagger_{-k\downarrow})+E_0
\end{align}
where $\Delta_k$ is the gap function giving by
\begin{align}\label{gap}
\Delta_k &= -{1\over N}\sum_{m} V_{m}\Delta_me^{-ik\cdot R_m}\\
E_0 &=  -{1\over N}\sum_{m} V_{m}|\Delta_m|^2
\end{align}
with $\Delta_m=<\sum_i c_{i\downarrow}c_{i+m\uparrow}>$.

An important feature is that the range of the coupling potential gives the symmetries of
the gap function, as listed in the Table I.

\text{Table I. Gap symmetries.}

\begin{tabular*}{8cm}{@{\extracolsep{\fill}}lllr}
 \hline\hline
m & condition &$\Delta_{k}$ & symmetry  \\
\hline
0 & & $\sim 1$ & s \\
$n.n.$ & (1)& $\sim(\cos k_x + \cos k_y)$ & $s^-$ \\
$n.n.$ & (2)& $\sim(\cos k_x - \cos k_y)$ & $d_{x^2-y^2}$ \\
$n.n.n.$ & (1)& $\sim\cos k_x \cos k_y$ &  $s_{x^2+y^2}$ \\
$n.n.n.$ & (2)& $\sim\sin k_x \sin k_y$ & $d_{xy}$ \\
\hline\hline
\end{tabular*}

In Table I $n.n.$ denotes the nearest neighbor sites, $x = (\pm 1, 0), y = (0, \pm 1)$,
condition (1) stands for the isotropic case $\Delta_x=\Delta_y$, and (2) for the
anisotropic case $\Delta_x =-\Delta_y$. In the case of the next nearest neighbor sites
(denoted by $n.n.n.$), $xy = (\pm 1, \pm 1), x\bar y = (\pm 1, \mp 1)$, condition (1)
stands for the isotropic  case $\Delta_{xy}=\Delta_{x\bar y}$, and (2) for the
anisotropic case $\Delta_{xy}=-\Delta_{x\bar y}$. In the above table the gap symmetry
$s_{x^2+y^2}$ has been found recently in the iron-based superconductors\cite{Mazin,Yu}.
Surprisingly, it is seen from the table that this neighbor coupling gives most symmetries
of gap functions that have been found in various superconductors, including conventional
and high-Tc superconductors with square lattices.

It should be emphasized that these gap symmetries have nothing to do with angular momenta
of the Cooper pairs. They originate from the point group symmetries of the crystals of
superconductors. In fact, strictly speaking, the angular momentum $\hat L$ of Cooper
pairs is not a conserved quantity since the potential does not obey the rotational
symmetries of group $R_3$\cite{yunping}, which is only true for the local atomic
orbitals. For the  neighbor coupling the point group symmetries are much more significant
than the atomic orbital symmetries.

 In order to confirm the isotropy and anisotropy of $\Delta_k$
Green's function  is defined as follows,
\begin{align}
<<c_{i\uparrow}(t)|c_{j\downarrow}(0)>>&=-i\theta(t)<[c_{i\uparrow}(t),c_{j\downarrow}(0)]_+>
\end{align}
 According to the standard method of Green's
function theory one obtains
\begin{align}
<<c_{k\uparrow}|c_{-k\downarrow}>>_\omega &= -{\Delta_k \over \omega^2 - \xi_k^2}, \xi_k
= \sqrt{\epsilon_k^2 + |\Delta_k|^2}
\end{align}
where $\epsilon_k$ are the energies of free electrons, $\epsilon_k= -2t(\cos k_x + \cos
k_y)+ 2t' \cos k_x\cos k_y$. Then the spectrum theorem of Green's function gives
\begin{align}
<c_{-k\downarrow}c_{k\uparrow}> &= -{1\over \pi} Im
<<c_{k\uparrow}|c_{-k\downarrow}>>_{\omega+i\eta}\nonumber \\
&= {\tanh(\xi_k/2k_BT)\over 2\xi_k}\Delta_k
\end{align}
Combining the above equation and (\ref{gap}) one obtains the gap equation
\begin{align}
 \Delta_k &=-\sum_{q} V({k-q})
{\tanh(\xi_q /2k_BT)\over 2\xi_q}\Delta_q\label{gap2}
\end{align}
where $V(q)$ are the fourier components given by $V(q) = (1/N)\sum_{m} V_m e^{-i{ q}\cdot
{ R}_m}$. Under the 4-fold rotations, $R$, of the point group symmetries of the potential
$U(m)$ this equation transforms in the following way(see Appendix),
\begin{align}\label{gap1}
 \Delta_{Rk} &=-\sum_{q} V({k-q}) {\tanh(\xi_q /2k_BT)\over 2\xi_q}\Delta_{Rq}
\end{align}
Compare the above equation with the gap equation (\ref{gap2}) one finds that solutions
with the following two symmetries exist:
\begin{align}\label{sym}
 \Delta_{Rk} &=\pm \Delta_k
\end{align}
Therefore, both conditions (1) and (2) in Table I can be realized, that is, both the
isotropic case and the anisotropic one are allowed. It is then confirmed that various gap
symmetries listed in Table I are allowed by the neighbor coupling. A real system selects
one of the symmetries with the lowest energy, or the largest gap magnitude.

The symmetries revealed in (\ref{sym}) are called respectively conventional and
unconventional superconductivity in literatures\cite{Annett}. In this paper it is proved
that both these two symmetries are in fact a conclusion of the 4-fold rotational
symmetries of the coupling potential on the square lattice superconductors. 

\begin{figure}
\label{tem}
\includegraphics[width=4cm,height=4cm]{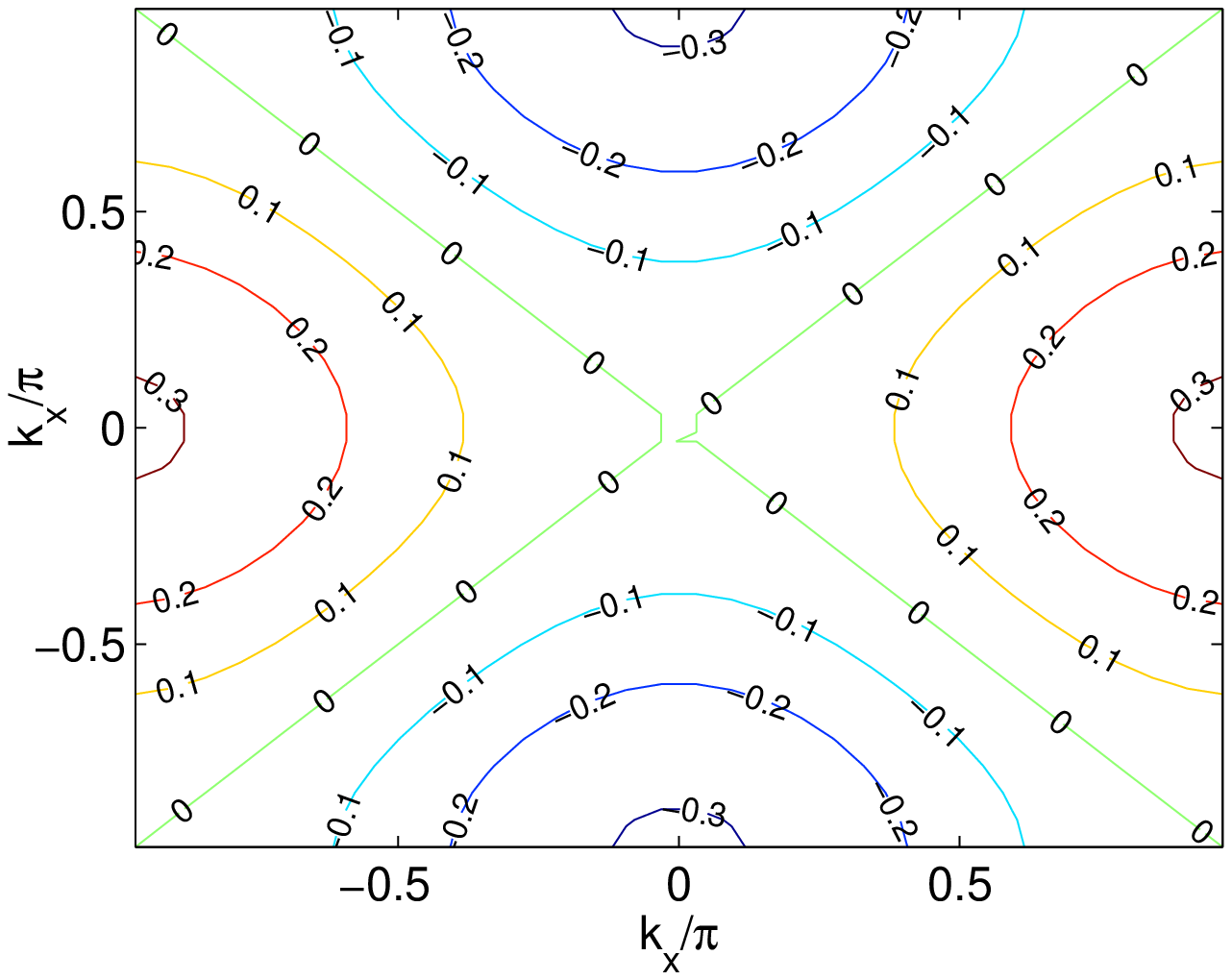}
\includegraphics[width=4cm,height=4cm]{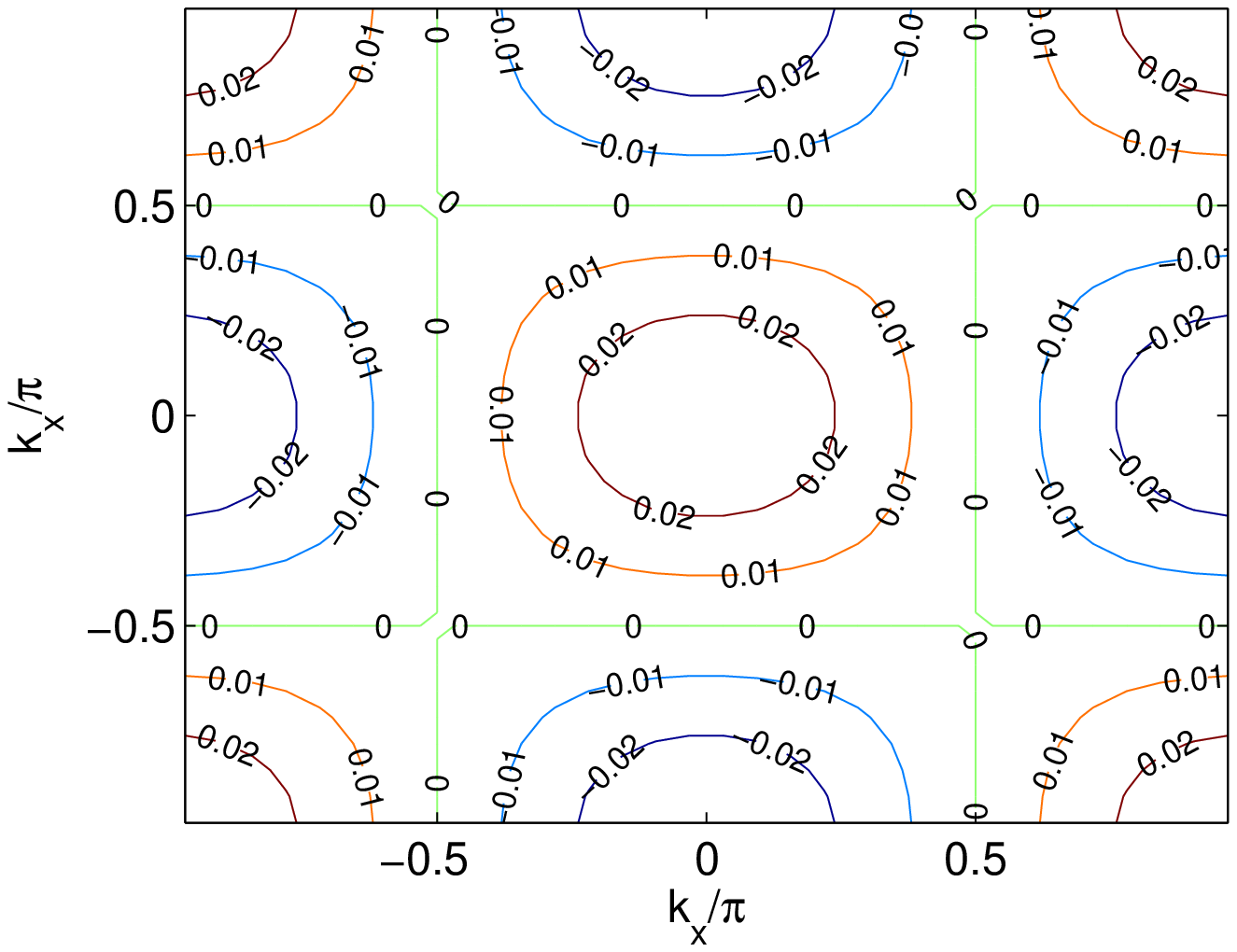}
\caption{Gap functions with the $d_{x^2 - y^2}$ symmetry due to the $n.n.$ coupling $t_1
= 1.0, t_2 = 0.0, U_0 = 0.0, U_1 = -1., U_2=0.0, \mu=0.0$(left panel) and with the
$s_{x^2y^2}$ symmetry due to the $n.n.n.$ coupling $t_1 = 1.0, t_2 = 0.0, U_0 = 0.0, U_1
= 0., U_2=-1.0, \mu=0.0$(right panel). }
\end{figure}

 The gap symmetries can be checked by numerical calculations. The gap equation
(\ref{gap2}) is solved self-consistently through iterations from a guessed initial gap
function. It is found that this equation has stable solutions for a wide range of
parameters. Two examples are shown in Fig.1  for the $d_{x^2 - y^2}$ and the $s_{x^2y^2}$
gap symmetries. With the same parameters one also finds $s^-$ and $d_{xy}$ gap symmetries
from a different  guessed initial gap function, but their magnitudes of gap functions are
a few orders smaller. That is to say, the two symmetries shown in Fig.1 are most
energetic preferential for the given parameters. Further computations show that different
guessed initial gap functions lead to  different gap symmetries, but the gap function
with the largest magnitude is almost determined by the neighbor coupling with lowest
potential value. Thus the earlier theoretical analysis of this paper is further
demonstrated by these calculations. It is concluded that the gap symmetries of cuprate
superconductivity are in fact due to the neighbor coupling with 4-fold rotational
symmetries of crystals of superconductors.

The gap symmetries provide information for the coupling range of Cooper pairs. Comparing
with the on-site coupling of the conventional $s$-wave BCS superconductivity the Cooper
pairs of high-Tc superconductors can distribute on two neighbors, including the $n.n.$
and the $n.n.n.$ neighbors. The size of the Cooper pairs is determined by the minimum of
the coupling potential. The neighbor coupling probably provides a unified
phenomenological model for the BCS and the high-Tc superconductivity.

\section{Conclusion}
In conclusion, 
it is found that the gap symmetries are simply determined by the 4-fold rotational
symmetries of the coupling potential on neighbor sites. A local on-site coupling
potential results in the on-site pairing with the conventional $s$-wave symmetry, but a
$n.n.$ or $n.n.n.$ coupling potential results in the pairing on neighbor sites with the
$s^-$, $d_{x^2-y^2}$, $s_{x^2y^2}$ or $d_{xy}$ symmetries. It is proved that both
isotropic and anisotropic gap functions are allowed by the 4-fold rotational symmetries
of the coupling potential. Finally a numerical computation is performed to demonstrate
the gap symmetries. The solutions of the gap equation are very stable for a wide range of
parameters.

 This work was supported by the National Natural Science Foundation of
China (Grant No. 10874049),  the State Key Program for Basic Research of China (No.
2007CB925204) and the Natural Science Foundation of Guangdong province ( No. 07005834 ).

\text{\bf Appendix}

Eq. (\ref{gap1}) can be proved in the following way. From (\ref{gap2}) one has
\begin{align}
 \Delta_{Rk}  &=-{1\over N}\sum_{qm} V_{m}e^{i(q-Rk)\cdot m}
{\tanh(\xi_q /2k_BT)\over 2\xi_q}\Delta_{q}\nonumber\\
&=-{1\over N}\sum_{qm} V_{m}e^{i(R^{-1}q-k)\cdot R^{-1}m}
{\tanh(\xi_q /2k_BT)\over 2\xi_q}\Delta_{q}\nonumber\\
&=-{1\over N}\sum_{qm} V_{Rm}e^{i(R^{-1}q-k)\cdot m}
{\tanh(\xi_q /2k_BT)\over 2\xi_q}\Delta_{q}\nonumber\\
&=-{1\over N}\sum_{qm} V_{m}e^{i(q-k)\cdot m} {\tanh(\xi_q /2k_BT)\over
2\xi_q}\Delta_{Rq}\nonumber\\&=-\sum_{q} V({k-q}) {\tanh(\xi_q /2k_BT)\over
2\xi_q}\Delta_{Rq}\nonumber
\end{align}
where the symmetry $\xi_{Rq}\approx \xi_q$ has been used since the value of $\xi_q$
depends weakly on $\Delta_q$ and $\epsilon_{Rq}=\epsilon_q$.

\end{document}